\documentclass[aip, apm, reprint, amsmath, amssymb, nolongbibliography, superscriptaddress]{revtex4-2}

\usepackage[T1]{fontenc}
\usepackage{palatino}
\usepackage{soul}
\soulregister\cite7
\soulregister\ref7
\soulregister\pageref7
\soulregister\ce7
\soulregister\onlinecite7
\usepackage{graphicx}
\usepackage[svgnames]{xcolor}
\usepackage{amsmath, bm}
\usepackage[detect-none]{siunitx}
\usepackage{fancyref}
\usepackage[colorlinks,
            linkcolor=DarkGreen,
            urlcolor=Blue,
            citecolor=DarkGreen]{hyperref}
\usepackage[utf8]{inputenc}
\usepackage{textcomp}
\usepackage[version=3]{mhchem}
\usepackage{miller}

\usepackage{pdfpages}
\usepackage{pgffor}
\usepackage{etoolbox}
\makeatletter
\patchcmd{\@outputpage@head}{\@ifx{\LS@rot\@undefined}{}{\LS@rot}}{}{}{}
\makeatother

\newcommand{\orcidicon}[1]{\href{https://orcid.org/#1}{\includegraphics[height=\fontcharht\font`\B]{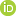}}}

\begin{document}

\title[]{Strain, Young's modulus, and structural transition of \ce{EuTiO3} thin films probed by micro-mechanical methods}

\author{Nicola~\surname{Manca}\,\orcidicon{0000-0002-7768-2500}}
\email{nicola.manca@spin.cnr.it}
\affiliation{CNR-SPIN, C.so F.\,M.~Perrone, 24, 16152 Genova, Italy}
\author{Gaia~\surname{Tarsi}\,\orcidicon{0009-0008-2685-8378}}
\affiliation{Dipartimento di Fisica, Università degli Studi di Genova, 16146 Genova, Italy}
\author{Alexei~\surname{Kalaboukhov}\,\orcidicon{0000-0003-2939-6187}}
\affiliation{Department of Microtechnology and Nanoscience – MC2, Chalmers University of Technology, SE 412 96, Gothenburg, Sweden}
\author{Francesco~\surname{Bisio}\,\orcidicon{0000-0003-1776-3023}}
\affiliation{CNR-SPIN, C.so F.\,M.~Perrone, 24, 16152 Genova, Italy}
\author{Federico~\surname{Caglieris}\,\orcidicon{0000-0002-9054-580X}}
\affiliation{CNR-SPIN, C.so F.\,M.~Perrone, 24, 16152 Genova, Italy}
\author{Floriana~\surname{Lombardi}\,\orcidicon{0000-0002-3478-3766}}
\affiliation{Department of Microtechnology and Nanoscience – MC2, Chalmers University of Technology, SE 412 96, Gothenburg, Sweden}
\author{Daniele~\surname{Marré}\,\orcidicon{0000-0002-6230-761X}}
\affiliation{Dipartimento di Fisica, Università degli Studi di Genova, 16146 Genova, Italy}
\affiliation{CNR-SPIN, C.so F.\,M.~Perrone, 24, 16152 Genova, Italy}
\author{Luca~\surname{Pellegrino}\,\orcidicon{0000-0003-2051-4837}}
\affiliation{CNR-SPIN, C.so F.\,M.~Perrone, 24, 16152 Genova, Italy}

\begin{abstract}
  \ce{EuTiO3} (ETO) is a well-known complex oxide mainly investigated for its magnetic properties and its incipient ferro-electricity. In this work, we demonstrate the realization of suspended micro-mechanical structures, such as cantilevers and micro-bridges, from 100\,nm-thick single-crystal epitaxial ETO films deposited on top of \ce{SrTiO3}(100) substrates. By combining profile analysis and resonance frequency measurements of these devices, we obtain the Young's modulus, strain, and strain gradients of the ETO thin films. Moreover, we investigate the ETO anti-ferro-distorsive transition by temperature-dependent characterizations, which show a non-monotonic and hysteretic mechanical response. Comparison between experimental and literature data allows us to weight the contribution from thermal expansion and softening to the tuning slope, while a full understanding of the origin of such a wide hysteresis is still missing. We also discuss the influence of oxygen vacancies on the reported mechanical properties by comparing stoichiometric and oxygen-deficient samples. \textbf{This is the author's peer reviewed, accepted manuscript. However, the online version of record will be different from this version once it has been copyedited and typeset. PLEASE CITE THIS ARTICLE AS DOI: 10.1063/5.0166762.}
\end{abstract}

\maketitle

\section*{Introduction}

\ce{EuTiO3} (ETO) is a complex oxide belonging to the titanate family. It is the closest compound to \ce{SrTiO3} (STO),\cite{Bussmann-Holder2011} which was extensively studied over the last decades and is among the standard substrate materials employed for the deposition of oxide thin films.
An interesting aspect of ETO is that it is iso-structural to STO, with almost identical lattice constants.\cite{Goian2012} This allows to grow very high quality thin films, having bulk-like characteristics, on top of STO substrates.\cite{Fujita2009}
At room temperature, \ce{EuTiO3} has Perovskite crystal structures with cubic lattice and, upon cooling, it undergoes a cubic to tetragonal transition at about 282\,K, driven by oxygen octahedra rotation,\cite{Kohler2012} which is similar to that observed in STO at 105\,K.\cite{Shirane1969}
Specific heat and thermal expansion measurements point toward a first-order phase transition, which also affects Young's modulus temperature dependence.\cite{Goian2012, Reuvekamp2014}
Thanks to its magnetic cation, ETO is also characterized by an anti-ferromagnetic transition at $T_{\mathrm{N}}$=5.5\,K.\cite{McGuire1966}
Moreover, it is an incipient ferroelectric and, despite its negative critical temperature of \textminus175\,K forbids the transition to a ferroelectric state, evidences of magneto-dielectric coupling made this compound an interesting candidate as multiferroic material.\cite{Katsufuji2001, Bussmann-Holder2015}

In recent years, the possibility to realize prototypical micro-electro-mechanical systems from complex oxides thin films has been demonstrated by taking advantage from selective chemical etching of different oxide compounds.\cite{Pellegrino2009, Manca2017, Manca2019a, Manca2020, Manca2021} In order to develop this new scientific and technological direction, it is of great interest to increase the number of viable oxide materials by discussing their fabrication protocols and characterizing their mechanical properties.
Contrary to STO, ETO is resistant to \ce{HF}, enabling selective chemical etching to realize suspended structures having desired shape.\cite{Plaza2021}

In this work, we investigate the mechanical properties of \ce{EuTiO3} by fabricating micro-mechanical structures from single crystal thin films deposited on top of STO(100).
ETO samples are initially characterized in terms of structural, magnetic, and optical properties and then micro-fabricated into suspended double-clamped bridges and cantilevers. The analysis of their mechanical properties allow us to quantify the ETO built-in strain, strain gradient along the in/out-of-plane directions, and Young's modulus.
Temperature-dependent measurements of the cantilever's resonance frequency allow us to investigate the ETO anti-ferro-distorsive transition by mechanical methods, which was analyzed taking into account contributions from thermal expansion and Young's modulus temperature dependence.
At last, we discuss how oxygen vacancies, one of the most common doping defect in complex oxides, affect the reported characteristics.

\section*{Experimental}

\ce{EuTiO3} thin films were grown by pulsed laser deposition on top of \ce{SrTiO3(100)} single-crystal substrates kept at 650\,$^\circ$C.  We employed a KrF excimer laser (248\,nm) with a repetition rate of 4\,Hz. The energy density on the target was 1.7\,J/cm$^\mathrm{2}$, and the distance between the target and the substrate was 50\,mm. The growth chamber base pressure was 1$\times$10$^{\mathrm{-6}}$\,mbar and, where not stated differently, the background oxygen pressure during the deposition was 1.5$\times$10$^{\mathrm{-4}}$\,mbar.
Thickness measurements are discussed in the Supplementary Material Sec.~I.
Atomic Force Microscope (AFM) imaging as has been performed by using the Bruker Dimension ICON AFM with Nanoscope 6 controller in tapping mode.
Suspended microstructures were fabricated by UV lithography of SPR-220 photo-resist spin-coated at 6000\,RPM for 45\,s, followed by baking at 120\,$^{\circ}$C for 135\,s.
ETO was removed by dry etching with an Argon milling system having sample water-cooling.
Etching time was 45\,mins and Ar ions energy was 500\,eV with a current density of 0.2\,mA/cm$^2$.
Samples cleaning from photo-resist residues required room-temperature ultrasonic baths of acetone followed by ethanol and then dried under nitrogen flow.
Selective etching of the STO substrate was obtained by put soaking the samples in a 5\,\% HF aqueous solution kept at 35\,$^{\circ}$C in bain-marie for 30\,mins.
During the bath, the samples were kept suspended above a magnetic stirrer rotating at 200\,RPM.
They are then transferred in a deionized water bath, followed by two different baths of pure ethanol to remove the presence of water.
Finally, they are dried in a \ce{CO2} critical point dryer system.
Mechanical characterizations were performed in a custom setup providing PID-controlled temperature and 2$\times$10$^{-5}$\,mbar base pressure. When not stated otherwise, the mechanical measurements were performed at the constant temperature of 25\,$^{\circ}$C.
All the mechanical spectra were recorded by measuring the thermal noise of the ETO cantilevers in the optical-lever detection scheme. We employed a 670\,nm laser focused on top of the structures with an optical power of 60\,\textmu W. The reflected light was converted into an electrical signal by a custom four quadrant photo-diode connected to a spectrum analyzer. The reported spectra were typically the result of 8 averages, with a bandwidth depending on the central frequency value starting from 1\,Hz when measuring around 13\,kHz.

\section*{Results}

\begin{figure}
  \includegraphics[width=\linewidth]{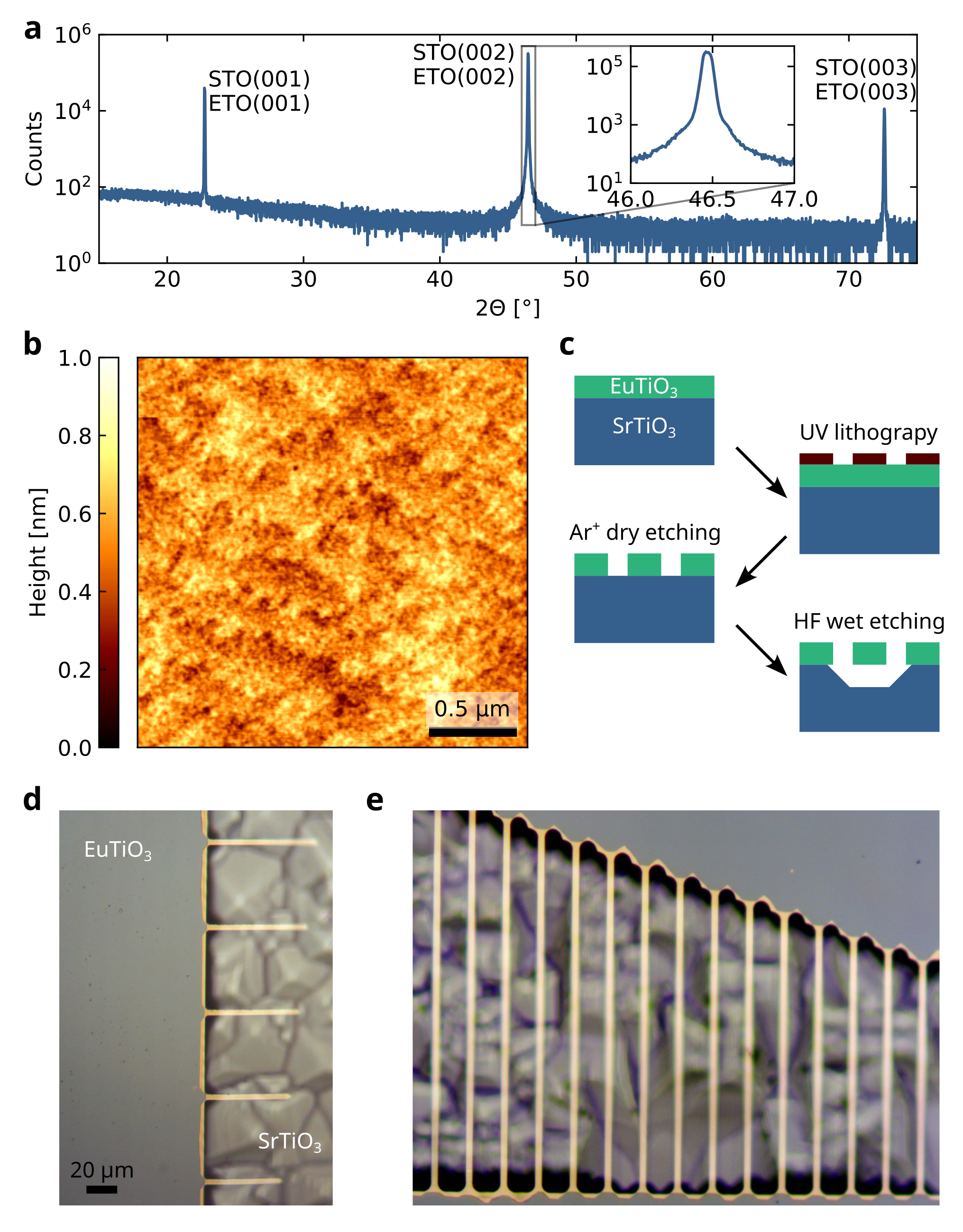}
  \caption{\label{fig:dev}
    \ce{EuTiO3} film characteristics and device fabrication.
    (a) XRD scan of a 100\,nm-thick ETO film. (00$l$) peaks are superimposed to the ones of the STO(001) substrate.
    (b) Tapping mode AFM topography image of the ETO surface.
    (c) Schematic fabrication steps of the ETO suspended structures. (d, e) Optical micrograph of ETO cantilevers and micro-bridges.
  }
\end{figure}

\begin{figure*}
  \includegraphics[width=\linewidth]{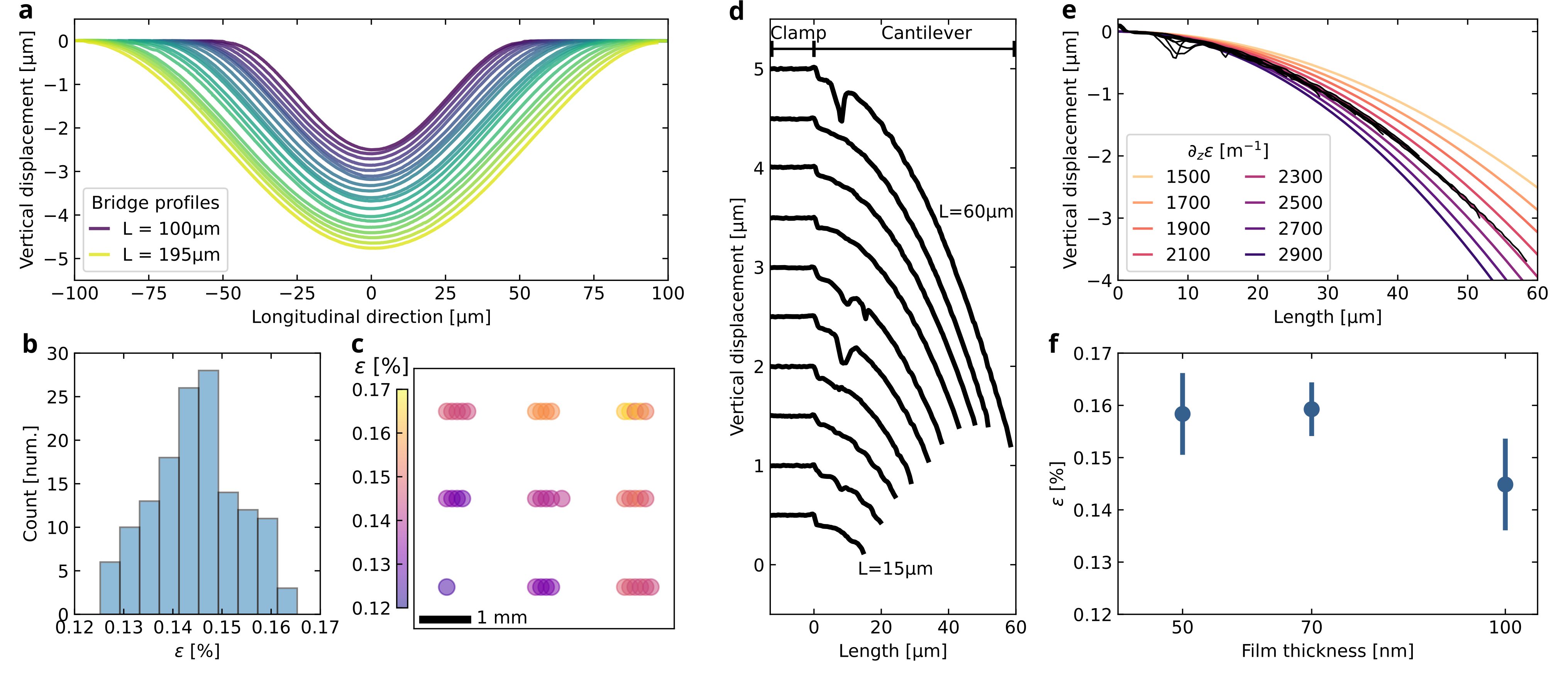}
  \caption{\label{fig:strain}
    Strain analysis of \ce{EuTiO_3} films.
    (a) Profiles of ETO double-clamped bridges having different length.
    (b) Histogram of the strain values calculated from the profile length of $\sim$140 buckled bridges.
    (c) Strain map of a ETO film, the square represents a 5$\times$5mm$^2$ substrate where each point is a bridge at its real position.
    (d) Profiles of cantilevers having different length showing the bending due to out-of-plane strain gradient. Vertical shift is 0.5\,\textmu m for better visibility.
    (e) Comparison between (black) superimposed experimental profiles from (d) and  (colors) simulated ones calculated for different strain gradient values.
    (f) Average strain values obtained for ETO films having different thickness, the bars indicate one standard deviation.
  }
\end{figure*}

The ETO crystal structure was investigated by X-ray diffraction, and a $\Theta$--$2\Theta$ scan of a 100\,nm-thick film is reported in Fig.~\ref{fig:dev}a.
ETO peaks cannot be resolved due to the superposition with those owning to the STO substrate, as further confirmed by reciprocal space maps reported in the Supplementary Material Sec.~II. This is in agreement with previous reports of stoichiometric bulk-like ETO films grown in similar conditions.\cite{Fujita2009}
Surface morphology was investigated by atomic force microscopy (AFM), showing very smooth ETO film surface with a RMS roughness of about 0.1\,nm over a scan area of 2.5$\times$2.5\,\textmu m$^2$, as shown in Fig.~\ref{fig:dev}b. This is representative of all the samples analyzed in this work, as reported in the Supplementary Material Sec.~III.
Our fabrication protocol is based on standard UV mask lithography and its main steps are schematically illustrated in Fig.~\ref{fig:dev}c, while details are discussed in the experimental section. ETO film is patterned by Ar ion milling and then cleaned in ultrasound acetone and ethanol baths,while selective etching of the STO substrate is obtained by put soaking the samples in HF diluted at 5\% in water.
The etching of STO starts out-of-plane, from the exposed regions, and then proceeds removing the substrate below the edges of the ETO film, making the narrower geometries suspended.\cite{Plaza2021}
In about 30 minutes all the structures having width below 5\,\textmu m are completely released and ready for the mechanical characterizations.

Examples of the two kind of geometries employed in this work, cantilevers and double-clamped bridges, are shown in Fig.~\ref{fig:dev}d and e, respectively, all having nominal width of 5\,\textmu m. In these pictures, clamped ETO is dark/blueish, while all the suspended regions are light/yellowish.
The typical pyramids that form on top of the STO(100) substrates after HF etching are visible on the background: these are the regions where ETO was removed by ion milling.
The nominal length of the cantilevers spans from 15 to 100\,\textmu m. In our samples all the cantilevers up to 60\,\textmu m were measurable, while all those above 75\,\textmu m were touching the substrate. This relatively sharp threshold is due to the balance between the downward bending of the cantilevers (signaling out-of-plane strain gradient) and the etching depth of the STO substrate, of about 6\,\textmu m, making the tip of longer ones to reach the substrate and sticking to that. Touching can be avoided by increasing the wet etching time, but in such a case liquid flow due to stirring may also become a limiting factor by bending and breaking longer structures. Double-clamped micro-bridges, instead, do not easily collapse and were fabricated with length from 100 to 250\,\textmu m. As visible in Fig.~\ref{fig:dev}e, their center is out-of-focus due to the relaxation of built-in compressive strain by buckling.

We can quantitatively evaluate the strain of an ETO film from the shape analysis of buckled double-clamped micro-bridges.\cite{Manca2021}
To do so, we measured the profile of each bridge by using an optical profilometer, which is an interferometric microscope providing an height map of its field of view.
An example of the profiles extracted from these maps is reported in Fig.~\ref{fig:strain}a, showing an array of ETO micro-bridges having length between 100 and 195\,\textmu m.
Notably, they are all bent downwards, which is the most common case in our samples. This could be related to the fabrication process or to strain relaxation at the clamping points. For each individual profile we calculated the best fit of a sum of trigonometric functions. The resulting analytical expression was then employed to obtain the profile length $L^{\mathrm{P}}$, which was compared to its nominal value $L$ to calculate the strain $\varepsilon$
\begin{equation}
  \label{eq:strain_def}
  \varepsilon = (L^{\mathrm{P}} - L) / L.
\end{equation}
This analysis is based on the assumption that all the in-plane compressive stress of the bridge is relaxed and converted into strain (elongation) upon the structure release.
The Python script implementing the strain analysis is included in the dataset associated to this work, as indicated in the ``Open Data'' section.

In this study we measured $\sim$140 micro-bridges fabricated on two different ETO thin film samples having thickness of of 97 and 100\,nm.
The calculated strain distribution of the whole dataset is reported in the histogram of Fig.~\ref{fig:strain}b, showing an average strain of $\overline\varepsilon$\,=\,+0.14\,\%\,$\pm$\,0.02\,\%, where the positive sign corresponds to compressive state.
The resulting $\overline\varepsilon$ implies an in-plane lattice compression of ETO films of $a_{\mathrm{STO}}\cdot\overline\epsilon/(1+\overline\epsilon)$\,=\,0.55\,pm. Under elastic deformation this would corresponds to an expansion of the $c$-axis in the out-of-plane direction of 0.85\,pm, calculated considering a Poisson's ratio of 0.22, as obtained from Ref.~\onlinecite{Li2015} (see Supplementary Material Sec.~IV).
Literature reports of ETO pseudo-cubic lattice constants at 300\,K indicate values between 3.860\,\AA\ and 3.908\,\AA.\cite{Fujita2009, Kohler2012, Li2015, Pappas2022}
Such dispersion is wider than our calculated lattice deformation due to epitaxial growth, making difficult to uniquely correlate the in-plane strain to out-of-plane lattice expansion from XRD data reported in Fig.~\ref{fig:dev}.
ETO films grown on top of STO are thus likely at the crossover between tensile and compressive strain, depending on the specific growth condition and crystal defects.
Our conclusion is that the measured compressive strain is the result of the formation of thermodynamically stable defects during the growth, such as oxygen vacancies or dislocations.

The width of the strain distribution was found to be related to long-range film inhomogeneities and not to random bridge-to-bridge variations.
This is shown in the strain map reported in Fig.~\ref{fig:strain}c, where the black frame represents the edges of the 5$\times$5\,mm$^2$ STO substrate. Each colored circle is located at the real position of a ETO micro-bridge and its color indicates the measured strain value. Strain magnitude increases from the bottom-left to the top-right corner, while bridges close to each other show smaller strain values dispersion.
Analysis of strain variation of bridges from the same set, such as those belonging to the same harp structure shown in Fig.~\ref{fig:strain}a, is reported in the Supplementary Material Sec.~V.
Similar characteristics were already observed in manganite thin films, and likely related to temperature gradients or to small variations of film stoichiometry due to substrate-plume misalignment during the growth.\cite{Manca2022}

While double-clamped bridges provide information about in-plane strain, cantilevers do not, because, thanks to their free end, stress is relaxed by elongating the structures and not by buckling. However, cantilevers can be employed to evaluate the strain gradient in the out-of-plane direction ($\partial_z\varepsilon_{x,y}$), because it results in vertical bending of the structures due to the different in-plane strain between top and bottom surfaces. Fig.~\ref{fig:strain}d shows a set of profiles owing to cantilevers having length between 15 and 60\,\textmu m. Here, the horizontal coordinate was shifted to align at $x$=0 the begin of the suspended regions, while all the profiles were vertically shifted of 0.5\,\textmu m for better visibility. The irregularities in the profiles are artifacts of the measurement technique, which fails to reconstruct the shape of these semi-transparent structures where the substrate is close below.
All the cantilevers are bent downwards, signaling a positive strain gradient in the vertical direction. Such gradient can be quantitatively evaluated by comparing the measured profiles with what expected from a simple finite element model based on a constant vertical gradient, which is reported in Fig.~\ref{fig:strain}e. Here, the black lines are all the measured profiles of \ref{fig:strain}d collapsed on top of each other, while the colored lines are profiles calculated for different values of $\partial_z \varepsilon$.
Best agreement is found for $\partial_z \varepsilon$=2300\,m$^{-1}$, with a confidence interval of about 5\,\%, i.e.\ one-half of the 200\,m$^{-1}$ line spacing. Since the ETO thickness is $t$=97\,nm, the resulting in-plane strain difference between the bottom and top surface is $\Delta \varepsilon$=$\partial \varepsilon_z\times t$=0.023\,\%, which is about 6 times smaller than the average in-plane strain.

To further understand the characteristics of strain and strain relaxation in ETO thin films, we grew films having different thicknesses and measured their average in-plane strain from the profile analysis of buckled micro-bridges. Apparently, thinner films have a slightly higher compressive strain, as can be seen in Fig.~\ref{fig:strain}f. From this comparison  we conclude that strain relaxation does not evolve layer-by-layer during the growth, otherwise we would observe an increase of strain value for thicker structures due to the positive gradient. It is instead a global characteristics of the crystal which likely evolves across its whole thickness as long as the growth process continues, at least for the explored film thickness range.

\begin{figure}
  \includegraphics[width=\linewidth]{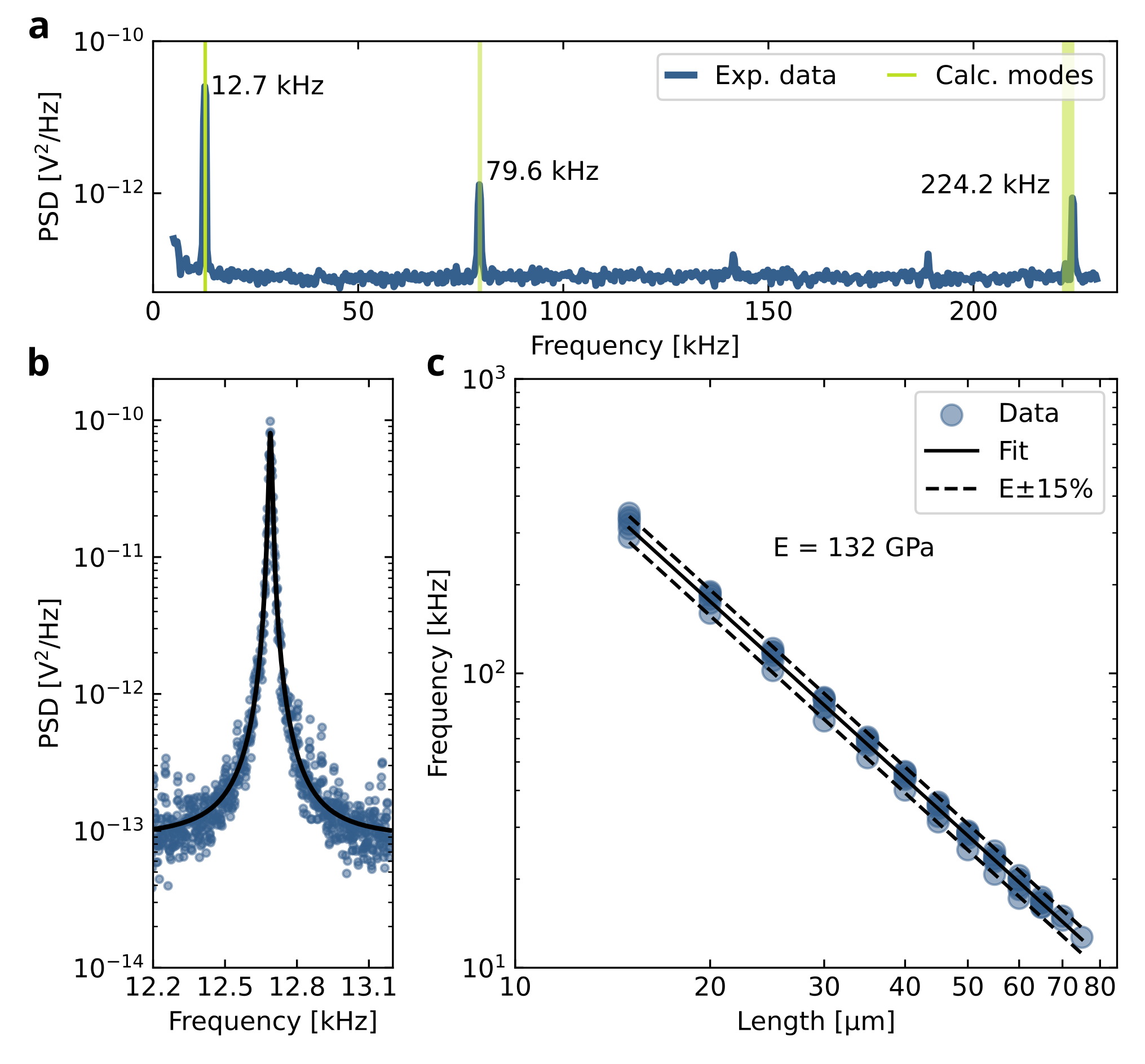}
  \caption{\label{fig:spectra}
    \ce{EuTiO_3} cantilever resonators.
    (a) Spectrum of a 75\,\textmu m-long cantilever. Green lines indicates the expected position of the flexural modes from Eq.~(\ref{eq:res_cant}) by rescaling the frequency of the first one. Line width indicates the uncertainty.
    (b) Thermal noise spectrum of the first flexural mode of a 75\,\textmu m-long ETO cantilever.
    (c) Frequency vs length for ETO cantilevers (blue dots), Young's modulus was obtained from fitting Eq.~(\ref{eq:res_cant}) (black solid line). All the data points lies is a $\pm$15\,\% range (black dashed lines) around the best fit value for $E$.
  }
\end{figure}

Even if slightly bent out-of-plane, micro-cantilevers can be employed as mechanical resonators to obtain the Young's modulus of the ETO film.\cite{Kiesewetter1992}
The resonance frequency can be calculated by using Euler-Bernoulli theory assuming a linear elastic material and small deflections. The resonance frequency of the flexural modes ($f_n$) of our cantilevers are modeled assuming thin and long beams having width much greater than their thickness (plate approximationq).\cite{Schmid2016}
\begin{equation}
  \label{eq:res_cant}
  f_n = \frac{\lambda^2_n}{2\pi}\frac{t}{L^2}\sqrt{\frac{E}{12\rho (1-\nu^2)}},
\end{equation}
where $\lambda_n$=1.8751, 4.6941, 7.8548, $(2n-1)\pi/2$ is a numerical parameter related to the mode shape, $t$ is the thickness, $L$ the length, $\rho$ the density, and $E$ the Young's modulus. We note that Eq.~(\ref{eq:res_cant}) has no stress dependence. This is because, as previously discussed, in cantilevers the longitudinal stress is relaxed due to their free end.
To confirm that the simple analytical model of Eq.~(\ref{eq:res_cant}) can be applied to our ETO resonators, we first measured a wide spectrum of a 75\,\textmu m-long cantilever, which is reported in Fig.~\ref{fig:spectra}a.
The first flexural mode is located at 12.7\,kHz, the second at 79.6\,kHz, and the third at 224.2\,kHz.
Since the mode spacing should be only given by the numerical factors $\lambda_n$ in Eq.~(\ref{eq:res_cant}), we can take the first mode as a reference
and calculate the expected values of the higher modes as $f_n = f_1\lambda_n/\lambda_1$.
The resulting frequencies are marked by the green bands in Fig.~\ref{fig:spectra}a and well-match the experimental values.
Fig.~\ref{fig:spectra}b shows a detailed spectrum of the first flexural mode, where the black line is a fit of the analytical expression of thermal noise spectrum.\cite{Hauer2013}. The resulting Q factor of 960 is in line with the other ETO cantilever resonators fabricated on these samples, all between 600 and 1300.
These values are much lower than what recently reported for other complex oxides micro-bridge resonators, exceeding 10k,\cite{Manca2022} this because double-clamped geometries under tension may take advantage from dissipation-dilution mechanism to enhance their Q factor.\cite{Sementilli2022}

We measured several ETO cantilevers having length spanning from 15 to 75\,\textmu m and reported their first eigenfrequency $f_1$ in the scatter plot of Fig.~\ref{fig:spectra}c.
In order to calculate the ETO Young's modulus ($E_{\mathrm{ETO}}$) from this dataset, we fit Eq.~(\ref{eq:res_cant}) considering a theoretical density of 6916.5\,kg/m$^3$, calculated from the \ce{Eu}, \ce{Ti}, and \ce{O} atomic masses and a (3.9\,\AA)$^3$ cubic unit cell volume.
To evaluate the dispersion of the measured resonance frequencies around their best fit  (black solid line), in Fig.~\ref{fig:spectra}c we mark the $E_{\mathrm{ETO}}\pm$15\,\% window (dashed black lines) comprising all the data points.
The resulting $E_{\mathrm{ETO}}$=132\,GPa is about one half of what obtained from measurements of resonant ultrasound spectroscopy in bulk single crystal samples, where the reported elastic moduli correspond to a Young's modulus of about 280\,GPa (see Supplementary Material Sec.~V).\cite{Li2015}
However, this result is in line with a recent work studying the mechanics of ultra-thin \ce{SrTiO3} which shows that the Young's modulus of 100\,nm-thick STO membranes is just below one half of the bulk value.\cite{Harbola2021}

\begin{figure}
  \includegraphics[width=\linewidth]{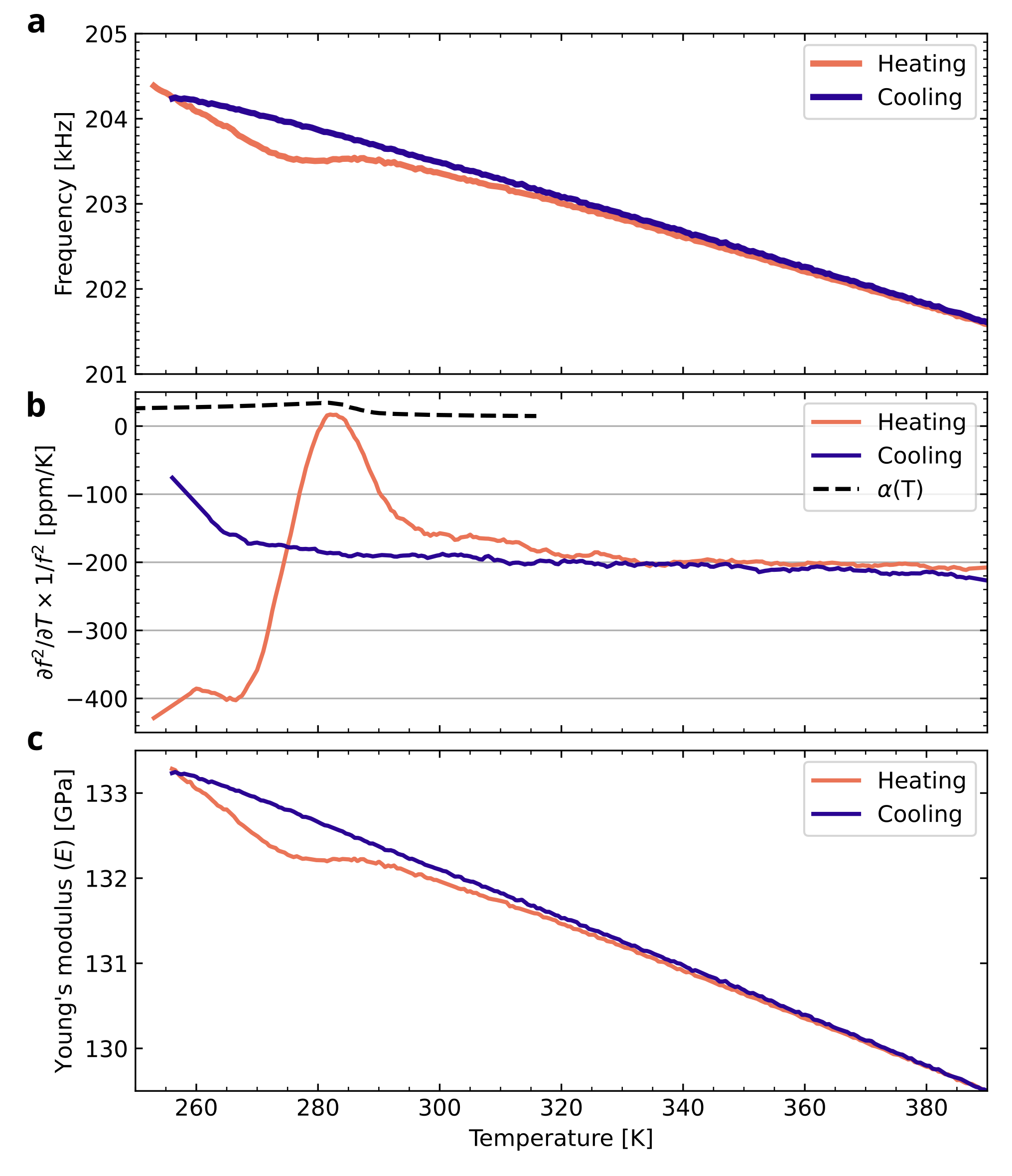}
  \caption{\label{fig:phase}
    Anti-ferro-distorsive transition of \ce{EuTiO_3} detected by resonance frequency measurements.
    (a) Frequency vs temperature characteristics of the first flexural mode of a 20\,\textmu m-long cantilever during heating and cooling ramps.
    (b) Relative temperature derivative of the squared frequency data from (a). Black dashed line is thermal expansion appearing in Eq.~(\ref{eq:freq2}), as extracted from Ref.\ \onlinecite{Goian2012}.
    (c) Temperature dependence of the ETO Young's modulus obtained from cantilever eigenfrequency measurements.
  }
\end{figure}

By measuring the temperature dependence of the resonance frequency of ETO cantilever resonators it is also possible to investigate the \ce{EuTiO3} anti-ferro-distorsive transition characteristics. To do so we measured a 20\,\textmu m-long cantilever which was initially cooled down from room temperature to 255\,K. The thermal noise spectrum of its first flexural mode was then recorded every 0.5\,K during both a heating and a cooling stepped ramps. The eigenfrequency vs temperature characteristics ($f_1(T)$) is reported in Figure~\ref{fig:phase}a, showing two distinct features: (i) a slope change during heating between 275\,K and 295\,K, (ii) a thermal hysteresis.
We can separate the contributions from thermal expansion and Young's modulus temperature dependence by considering the relative temperature derivative of the squared frequency
\begin{align}
  \label{eq:freq2}
  \frac{\partial_T f^2_1}{f^2_1} = \alpha(T) + \frac{\partial_T E}{E},
\end{align}
where $\alpha(T)$ is the linear expansion coefficient. Eq.~(\ref{eq:freq2}) was obtained from Eq.~(\ref{eq:res_cant}) as discussed in the Supplementary Material, Sec.~VI.
In Fig.~\ref{fig:phase}b we compare the experimental $\partial_T f^2_1 / f^2_1$ (solid lines) with the thermal expansion coefficient as exacted from Ref.\ \onlinecite{Reuvekamp2014} (dashed line).
Thermal expansion is small if compared to the experimental measurements, peaking at 35\,ppm/K at 282\,K and flattening out at about 15 ppm/K at high temperatures. As a consequence, Young's modulus provides the dominant contribution to the measured slope, with a high-temperature constant value of about $-$200\,ppm/K.
From the $f_1(T)$ data we can also obtain the Young's modulus of ETO by considering the first-order thermal expansion approximation for $t$, $L$, and $\rho$ in Eq.~(\ref{eq:res_cant}).
\begin{equation}
  \label{eq:freq2_E}
  E(T) = E_{T_0} \frac{f^2_1}{f^2_{1, T_0}}\frac{1}{(1 + A(T))}
\end{equation}
where the $T_0$ is reference temperature of 298\,K at which we measured the Young's modulus of 132\,GPa (see Fig.~\ref{fig:spectra}c) and $A(T)$ is the integral function of the thermal expansion $\alpha(T)$ calculated bewteen $T_0$ and $T$, as discussed in the Supplementary Material, Sec.~VI.

The anomaly of Young's modulus, marked by the oscillatory behavior of $\partial_T f^2_1 / f^2_1$ below 300\,K, indicates the onset of the anti-ferro-distorsive transition. This is in agreement with a previous report from J.~Schiemer and coworkers,\cite{Schiemer2016} showing large variations of the stiffness moduli in the proximity of the transition temperature in both single crystal and policristalline bulk ETO samples measured by resonant ultrasound spectroscopy.
  A key difference between our results and these previous studies is the width of the thermal hysteresis, that in their experiment was of about 1\,K. In our case, instead, thermal hysteresis is of about 35\,K, as estimated form Fig.~\ref{fig:phase}b.
  This wider transition could be the due to the high surface/volume ratio of our thin film-based devices, since hysteresis widening driven by size effect has been already reported for other materials showing first order phase transition.\cite{Brassard2005, Nazari2013}
  Another possible contribution is the presence of thermal gradients along the cantilever, since clamping point is in better contact with the thermal bath and the tip dissipates more.
However, considering the background pressure of 2$\times$10$^{-5}$\,mbar, such effect is expected to be small.
Because of the limited temperature span of our measurement setup, we could not access the whole hysteresis width.
Such hysteretic behavior was observed in several devices fabricated on different samples, as reported in Supplementary Material, Sec.~VII, where differences in hysteresis width and amplitude are related to the nature of nested loops, whose shape is critically dependent from the extremal temperatures.
Detailed analysis of this behavior will require dedicated experiments involving wider temperature range, samples having different thickness and cantilever geometries, and, ideally, magnetic field.

\begin{figure}
  \includegraphics[width=\linewidth]{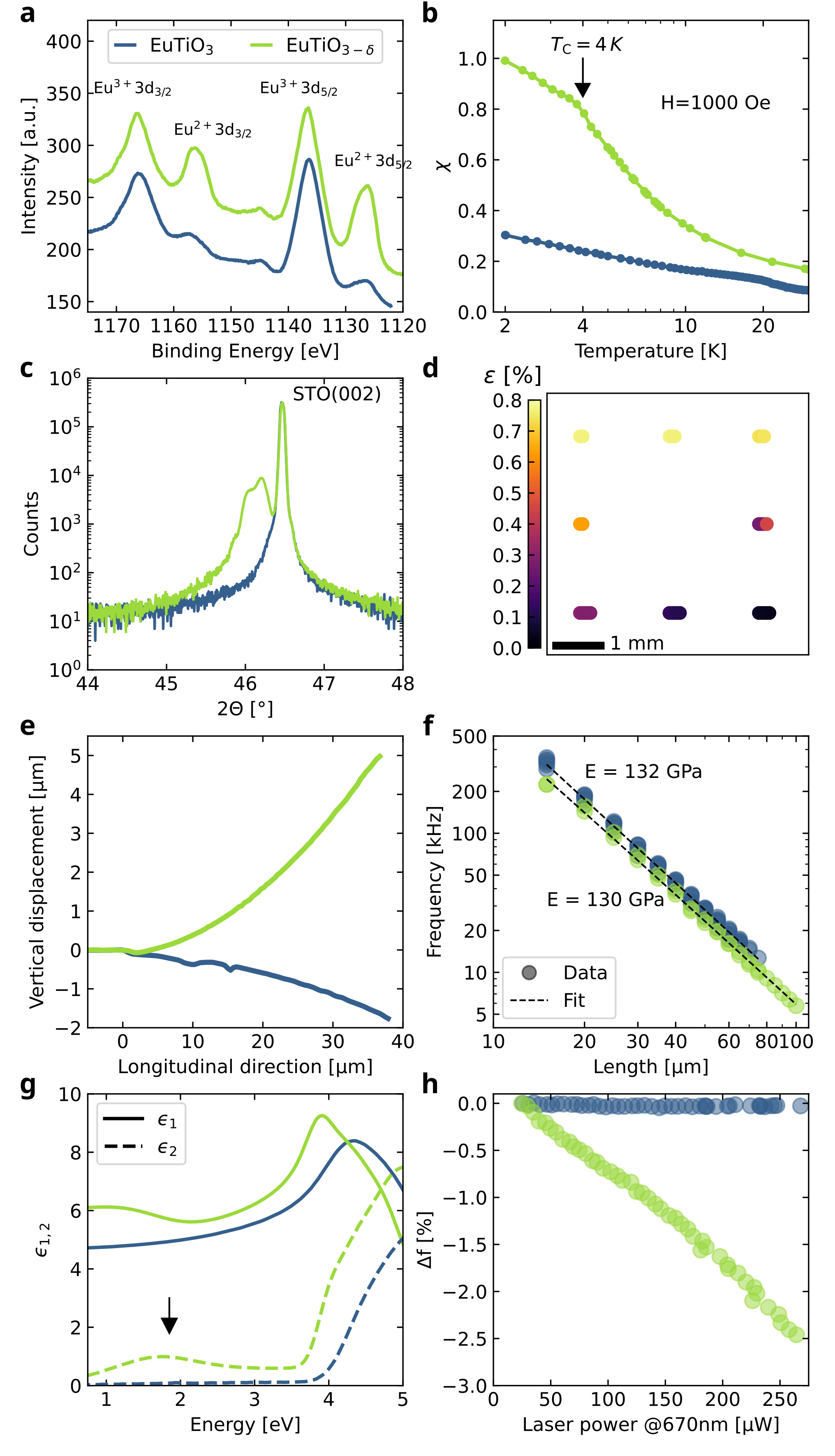}
  \caption{\label{fig:vacs}
    Comparison between \ce{EuTiO_3} (blue) and \ce{EuTiO_{3-\delta}} (green).
    (a) XPS spectra of the different Eu oxidation states.
    (b) SQUID measurement showing paramagnetic response and ferro-magnetic transition (arrow) \ce{ETO_{3-\delta}}.
    (c) XRD around the (002) peak of the STO substrate.
    (d) Strain map of a \ce{ETO_{3-\delta}} sample calculated as in Fig.~\ref{fig:strain}c.
    (e) Profiles of 40\,\textmu m cantilevers signaling opposite out-of-plane strain gradients.
    (f) Young's modulus obtained from the resonance frequency of ETO cantilevers. Data of stoichiometric sample are from Fig.~\ref{fig:spectra}c. 
    (g) Dielectric constants of ETO thin films. The black arrow marks the wavelength of the laser employed in (h).
    (h) Frequency shift vs optical power measured on two 65\,\textmu m-long cantilevers.
  }
\end{figure}

Finally, we investigated how the \ce{EuTiO3} characteristics reported so far are affected by oxygen vacancies, which are a typical defect as well as doping mechanism of complex oxides. To do so, we grew a nominally 100\,nm-thick films at the lower oxygen pressure of 10$^{-6}$\,mbar instead of 10$^{-4}$\,mbar. Its actual thickness, however, was 83\,nm, as reported in the Supplementary Material, Sec.~I.
After the preliminary characterizations reported in Fig~\ref{fig:vacs}a--c, \ce{ETO_{3-\delta}} samples underwent the same fabrication protocol of stoichiometric ones.
The formation of oxygen vacancies is indicated by the XPS data reported in Figure\ \ref{fig:vacs}a, where the valence change of Eu atoms, with added weight on \ce{Eu^{2+}}, is a consequence of charge balance after oxygen removal. Another feature of \ce{ETO_{3-\delta}} films is the onset of a ferro-magnetic transition at low temperature evidenced by an inflection point in the $\chi(T)$ characteristic.\cite{Fujita2009} This feature is visible in our SQUID data as evidenced by the black arrow in Fig.~\ref{fig:vacs}b.
XRD measurements reported in Fig.~\ref{fig:vacs}c shows that oxygen-deficient samples have larger out-of-plane inter-planar distance. Such  lattice expansion is a common feature of oxygen vacancies-doped complex oxides,\cite{Choi2009, Mattoni2018} which is associated to an increase in the compressive strain of the material. Moreover, the broadening of the diffraction peak is quite large, also showing peak splitting.

A peculiar characteristics of \ce{ETO_{3-\delta}} samples is strain inhomogeneity, as exemplified by the strain map of Fig.~\ref{fig:vacs}e, which shows strong variations between the upper and the lower edges. Similar strain gradients were observed across four samples grown in different deposition runs and always associated to broad XRD peaks, similar to what reported in Fig.~\ref{fig:vacs}c.
Because of that, we hypothesize that oxygen deficiency during growth makes ETO much more sensitive to small temperature gradients or substrate inhomogeneities, which, in higher pressure condition, do not affect long-range strain characteristics.
Another dramatic difference between stoichiometric and vacancy-doped cases is the out-of-plane strain gradient.
Fig.~\ref{fig:vacs}f compares the profiles of two 40\,\textmu m-long cantilevers: the \ce{ETO_{3-\delta}} one is bent upwards and shows a much larger displacement than the stoichiometric case. Strain gradient is thus negative and about three times larger.
This feature is quite surprising, because in a simple layer-by-layer growth picture we could expect a better film oxidation at the bottom layers due to oxygen exchange with the STO substrate. However, in such a case the bottom surface of the cantilever would be more compressed, resulting in a profile shape similar to the \ce{ETO3} case. This result supports our previous conclusion that strain and strain gradients behave as a global film characteristics determined during the entire deposition process.
We also investigate whether the Young's modulus of \ce{ETO_{3-\delta}} was affected by oxygen vacancies by analyzing the frequency vs length relationship of an array of cantilevers, as in Fig.~\ref{fig:spectra}c.
As reported in Fig.~\ref{fig:vacs}f, this is not the case, beacuse the difference between the resulting Young's moduli, of just 2\,GPa, is within the experimental error, while the lower  resonance frequencies for the \ce{ETO_{3-\delta}} cantilevers are due to the thickness of the films ($t_{\mathrm{{ETO_3}}}$=97\,nm and $t_{\mathrm{ETO_{3-\delta}}}$=83\,nm).

It is well known that the formation of oxygen vacancies affects the light absorption of complex oxides.\cite{Blanc1971} To understand how this happens in \ce{ETO_{3-\delta}}, we assessed the optical properties of stoichiometric and oxygen-deficient films, prior to microfabrication, by means of spectroscopic ellipsometry. The complex dielectric functions $\epsilon$ of the substrate and the ETO film were modeled as a superposition of PSEMI oscillators (parametrized functions used to describe the optical response of crystalline semiconductors).\cite{Johs1998, Sygletou2023} The values of $\epsilon$ extracted from the ellipsometry data, reported in Fig.~\ref{fig:vacs}g, indicate that stoichiometric samples show no absorption below the band-gap, located at about 4\,eV, in agreement with previous reports.\cite{Stuhlhofer2016}
Oxygen vacancies determine a small redshifting of the band gap, from 4.0\,eV to 3.6\,eV, but critically increase $\epsilon_2$ (green dashed line) in the low-energy region, signaling the formation of in-gap states leading to broadband light absorption.
An increased optical absorption affects the response of mechanical resonators to incident light. We thus compared the resonance frequency shift of two 65\,\textmu m-long cantilevers, one made of \ce{ETO3} and the other from \ce{ETO_{3-\delta}} as a function of the incident light power. The laser is the same employed to probe the motion of the structure and its wavelength of 670\,nm corresponds to the arrow in Fig.~\ref{fig:vacs}g. The resulting tuning slopes are reported in Fig.~\ref{fig:vacs}h, where, for better comparison, we considered the relative frequency shift with respect to the lowest power employed in the experiment (25\,\textmu W). The stoichiometric sample shows almost no softening, and thus no heating, in agreement with its good transparency. \ce{ETO_{3-\delta}} cantilever, instead, lowers its eigenfrequency of about 2.5\,\% for an incident light power of $P$=250\,\textmu W. This can be compared to the frequency shift measured during the temperature ramps in Fig.~\ref{fig:phase}b by considering the finite differences
\begin{align}
  \label{eq:pow}
  \frac{\Delta f^2}{\Delta P}\frac{1}{f_0^2} =
    \frac{(1.025f_0)^2-f_0^2}{f_0^2 \Delta P} =
    \frac{0.05}{\Delta P} = 200 \left[ \frac{ppm}{\mu W} \right].
\end{align}
Quite nicely, we found a 1\,\textmu W $\leftrightarrow$ 1\,K equivalence, which holds for this specific cantilever length and above 300\,K.
 
\section*{Conclusions}

In summary, we investigated the mechanical properties of stoichiometric and oxygen vacancy-doped \ce{EuTiO3} by fabricating suspended micro-structures from single-crystal thin films grown on top of \ce{SrTiO3(001)} substrates.
Mechanical characterization measurements provided a quantitative evaluation of average in/out-of-plane strain and strain gradients as well as the Young's modulus of the material.
Stoichiometric films are found slightly compressed, with an average strain of +0.14,\%, while oxygen-deficient ones shows higher strain inhomogeneity.
The Young's modulus of about 132\,GPa is almost one-half of what found in bulk samples, in line with recent measurements on \ce{SrTiO3} membranes having similar thickness.
Temperature-dependent mechanical measurements indicates that above 300\,K the relative derivative of the Young's modulus is about $-$200\,ppm/K.
Below 300\,K we observe a non-monotonic and hysteretic mechanical response associated to the \ce{EuTiO3} anti-ferro-distorsive transition. Its width of about 35\,K is in striking difference with respect to bulk samples that show a small histeresis of about 1\,K. Such wideding could be related to size effect or temperature gradients across the structure, but a clear understanding of its origin will require new dedicated experiments.

\section*{Supplementary Material}

The Supplementary Material of this article contains \ce{EuTiO3} growth data and thickness calibration, reciprocal space maps of stoichiometric and oxygen-deficient samples, surface topography of as-grown \ce{EuTiO3} thin films, evaluation of bulk \ce{EuTiO3} Young's modulus and Poisson's ratio from literature data, analysis of strain variations within bridges owning to the same harp array, derivation of Eq.~(\ref{eq:freq2}) and Eq.~(\ref{eq:freq2_E}), and frequency vs temperature data, as those reported in Fig.~\ref{fig:phase}a, measured on different cantilevers.

\section*{Acknowledgments}

We thank Emilio Bellingeri, Gianrico Lamura,  Cristina Bernini, and Alejandro Enrique Plaza for the useful discussions.
This work was carried out under the OXiNEMS project (\href{www.oxinems.eu}{www.oxinems.eu}). This project has received funding from the European Union’s Horizon 2020 research and innovation programme under Grant Agreement No. 828784. We acknowledge financial support from the Universit\`a di Genova through the ``Fondi di Ricerca di Ateneo'' (FRA). We also acknowledge support from the Swedish infrastructure for micro- and nano-fabrication - MyFab.

\section*{Open Data}

The numerical data shown in the figures of this manuscript and the
supplemental material can be donwloaded from the Zenodo online
repository: ~\href{http://dx.doi.org/10.5281/zenodo.8109185}{http://dx.doi.org/10.5281/zenodo.8109185}

\bibliography{Library.bib}


\newpage\newpage



\section*{Supplementary material}

## Supplementary Material

—

# Strain, Young's modulus, and structural transition of EuTiO$_3$ thin films probed by micro-mechanical methods

Nicola Manca 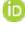,[1, *] Gaia Tarsi 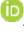,[2] Alexei Kalaboukhov 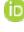,[3] Francesco Bisio 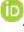,[1]

Federico Caglieris 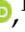,[1] Floriana Lombardi 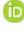,[3] Daniele Marré 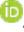,[2, 1] and Luca Pellegrino 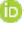[1]

$^1$*CNR-SPIN, C.so F. M. Perrone, 24, 16152 Genova, Italy*

$^2$*Dipartimento di Fisica, Università degli Studi di Genova, 16146 Genova, Italy*

$^3$*Department of Microtechnology and Nanoscience — MC2,*

*Chalmers University of Technology, SE 412 96, Gothenburg, Sweden*

This supplemental material contains the following:

- Section I: EuTiO$_3$ growth data and thickness calibration.

- Section II: Reciprocal space maps of stoichiometric and oxygen-deficient samples.

- Section III: Surface topography of as-grown EuTiO$_3$ thin films.

- Section IV: Evaluation of bulk EuTiO$_3$ Young's modulus and Poisson's ratio.

- Section V: Bridge-to-bridge strain variations.

- Section VI: Derivation of frequency vs temperature analysis equations.

- Section VII: Frequency vs temperature data measured on different cantilevers.

———

\* nicola.manca@spin.cnr.it

## I. EuTiO$_3$ GROWTH DATA AND THICKNESS CALIBRATION

| Sample | Growth Pressure | Laser Pulses | Estimated Thickness | Measured Thickness |
|--------|-----------------|--------------|---------------------|--------------------|
|        | [mbar]          | [num.]       | [nm]                | [nm]               |
| C2201  | $1.5 \cdot 10^{-4}$ | 1000     | 52.5*               | $50 \pm 1.3$       |
| C2202  | $1.5 \cdot 10^{-4}$ | 1500     | 78.8                | $71 \pm 2$         |
| C2203  | $1.5 \cdot 10^{-4}$ | 2000     | 105                 | $100 \pm 2.5$      |
| C2205  | $5.0 \cdot 10^{-6}$ | 2000     | 105                 | -                  |
| C2206  | $5.0 \cdot 10^{-6}$ | 2000     | 105                 | $83 \pm 2$         |
| C2217  | $1.0 \cdot 10^{-4}$ | 2000     | 105                 | $97 \pm 5$         |

TABLE S1. Growth and thickness data of the EuTiO$_3$ thin films employed in this work. Nominal thickness value was evaluated from X-ray reflectivity (XRR) measurements on C2201, here are marked by an asterisk, while thicker samples were grown by rescaling the number of laser pulses. XRR data for C2201 are reported in Figure S1. Thickness values were then measured for selected film by spectroscopic ellipsometry, as reported in Figure S2.

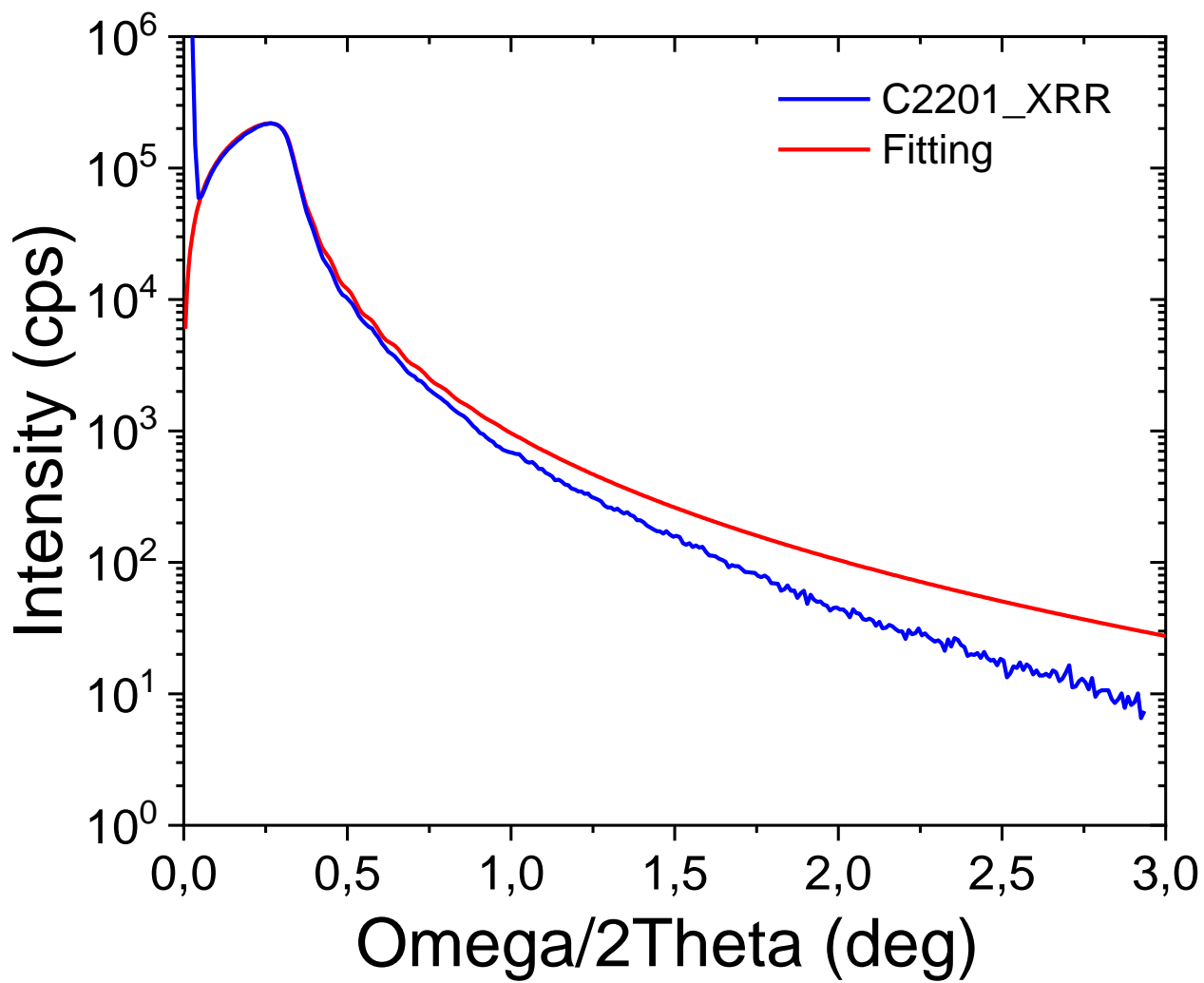

Figure S1.  X-ray reflectivity measurement performed on C2201.  Thickness value obtained by fitting intensity data is 52.5 nm and was employed as reference growth rate (nm per laser pulse) for all the other samples.

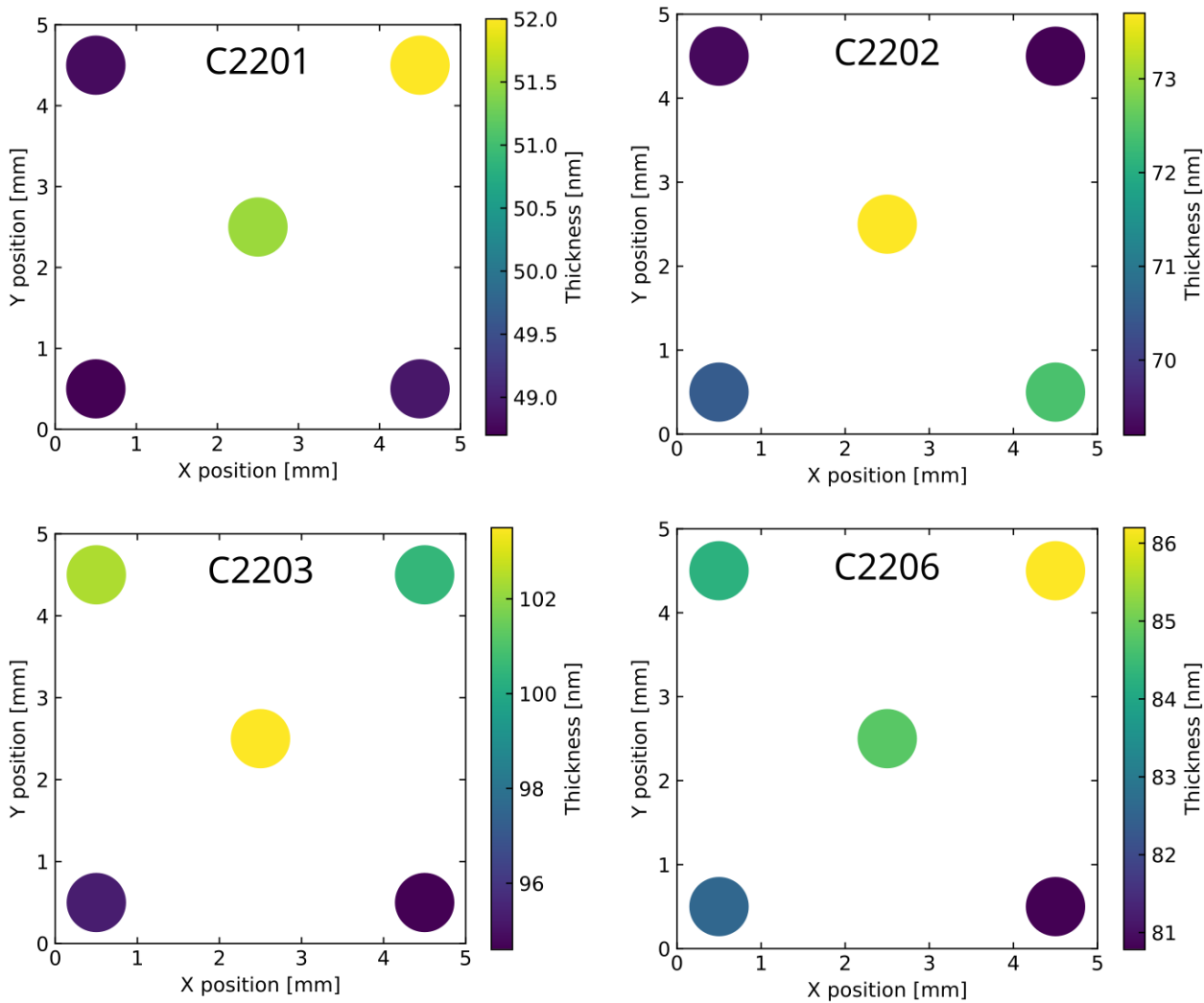

Figure S2. Thickness values obtained from spectroscopic ellipsometry measurements. Optical spectra were acquired at each corner and in the center of the sample. Spot size was about $1\,\text{mm}^2$. Film thickness reported in Table S1 is the average value from the five measurements. C2217 was measured by acquiring a single spectrum over a larger area covering almost all the sample surface.

## II. RECIPROCAL SPACE MAPS OF STOICHIOMETRIC AND OXYGEN-DEFICIENT SAMPLES

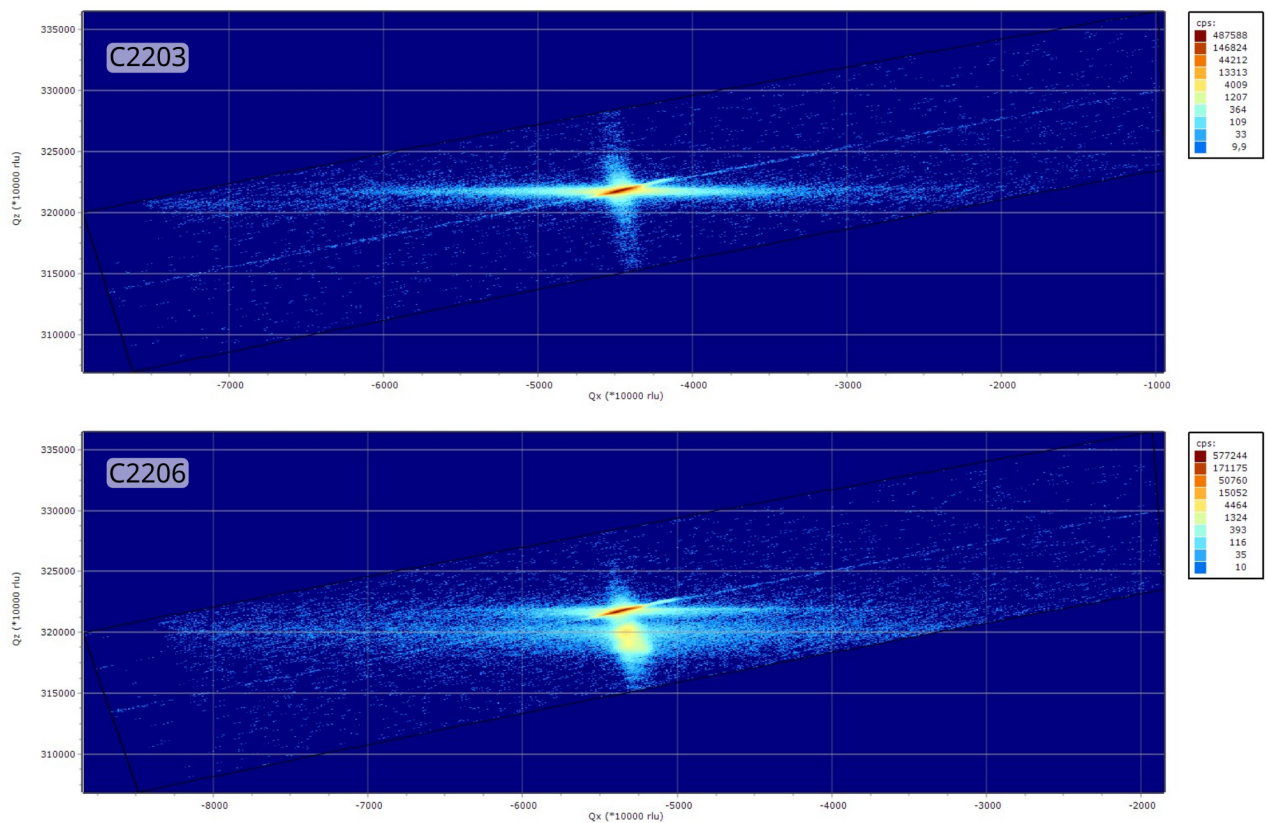

Figure S3. Reciprocal space maps around STO (002) diffraction peak for stoichiometric (C2203, upper panel) and oxygen deficient (C2206, bottom panel) samples.

## III. SURFACE TOPOGRAPHY OF AS-GROWN EuTiO3 THIN FILMS

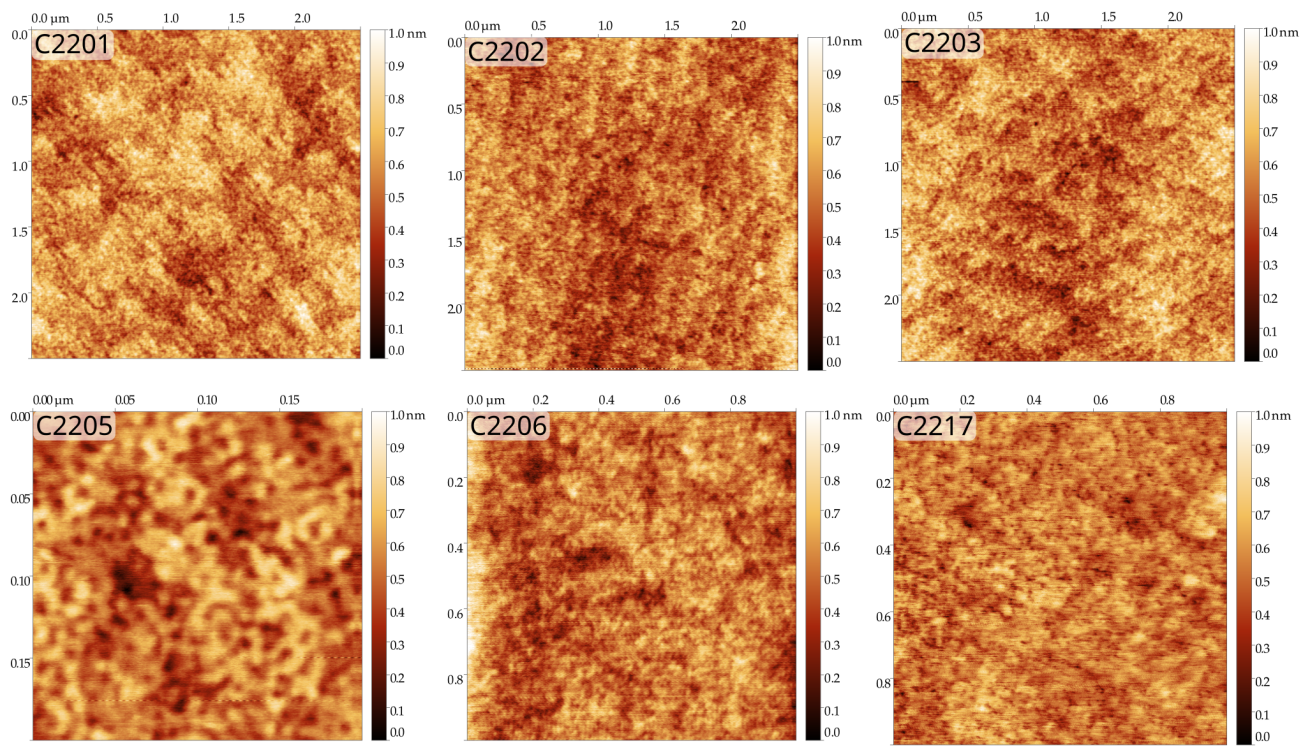

Figure S4. Tapping mode AFM Surface topography images of the EuTiO$_3$ thin films employed in this work. The vertical range is the same for all the panels. Images were flattened by line-by-line first-order polynomial subtraction.

| Sample Name | RMS roughness [nm] |
| :---: | :---: |
| C2201 | 0.12 nm |
| C2202 | 0.13 nm |
| C2203 | 0.13 nm |
| C2205 | 0.13 nm |
| C2206 | 0.13 nm |
| C2217 | 0.10 nm |

TABLE S2. Roughness of the EuTiO$_3$ thin films employed in this work calculated from the root-mean-square (RMS) of the height values measured by atomic force microscopy and reported in Figure S4.

## IV. EVALUATION OF BULK EuTiO₃ YOUNG'S MODULUS AND POISSON'S RATIO

For a linear elastic material, the stress–strain relationship is modeled by the fourth rank stiffness tensor $C_{ijkl}$:

$$\sigma_{ij} = C_{ijkl} \varepsilon_{kl} \tag{SE 1}$$

where $\sigma_{ij}$ is the stress tensor and $\varepsilon_{kl}$ the strain tensors. For isotropic linear elastic and homogeneous materials, the components of the stiffness tensor can be rewritten as

$$c_{ijkl} = \lambda \delta_{ij} \delta_{kl} + \mu (\delta_{ik} \delta_{jl} + \delta_{il} \delta_{jk}) \tag{SE 2}$$

where $\delta_{ij}$ is the Kronecker delta while $\lambda$ and $\mu$ are the Lamé constants. Lamé constants are related to engineering constants (Young's modulus $E$, Poisson's ratio $\nu$, and shear modulus $G$) as follows:

$$E = \frac{\mu(2\mu + 3\lambda)}{\mu + \lambda} \; ; \; \nu = \frac{\lambda}{2(\mu + \lambda)} \; ; \; \mu = G = \frac{E}{2(1 + \nu)} \tag{SE 3}$$

We can thus obtain $E$, $\nu$, and $G$ from the components of the stiffness tensor and Eq. (SE 2) and (SE 3). The components of $C$ for bulk single-crystal EuTiO₃ have been measured by resonant ultrasound spectroscopy as reported in L. Li et al., Phys. Rev. B 92, 024109 (2015). In the table below we show the data which are relevant in our analysis.

| Material | Temperature [K] | Theoretical Density [g/cm³] | Density [g/cm³] | $C_{11}$ [GPa] | $C_{44}$ [GPa] | $C'$ [GPa] | Error % |
|----------|-----------------|------------------------------|------------------|-----------------|-----------------|-------------|---------|
| EuTiO₃ | 288 | 6.914 | 6.744 | 307.24 | 116.30 | 100.06 | 0.2833 |

TABLE S3. EuTiO₃ material characteristics and elastic moduli in Voigt notation from L. Li et al., Phys. Rev. B 92, 024109 (2015).

Measured elastic moduli are related to Lamé constants by Eq. (SE 2) as follows

$$C_{11} = \lambda + 2\mu \; ; \; C_{44} = \mu \; ; \; C' = \frac{1}{2}(C_{11} - C_{12}) \rightarrow C_{12} = C_{11} - 2C' = \lambda \tag{SE 4}$$

These relationship are redundant, since three measured values provide two constants. While $\mu$ is directly related to a measured quantity, $\lambda$ can be obtained from $C_{11}$ or $C_{12}$, giving 74.64 GPa and 107.12 GPa, respectively. We thus have an uncertainty, although relatively small, in the evaluation of $E$ (278 – 288 GPa) and $\nu$ (0.20 – 0.24).

## V. BRIDGE-TO-BRIDGE STRAIN VARIATIONS

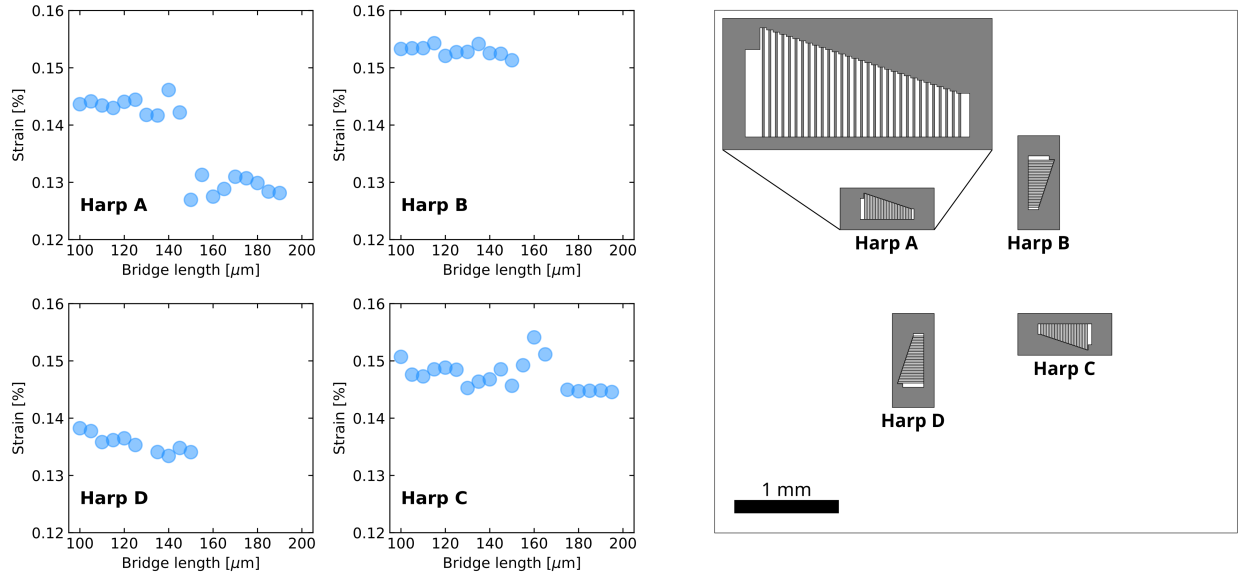

Figure S5. Strain measured on each bridge of four harp blocks. The device design is reported in the right half of the figure, showing the absolute position of each bridge with respect to the substrate (black squared frame). All the bridges owning to the same block are clamped to the common frame, whose inner white part is etched to make the bridges suspended. These data were taken on the same device shown in Fig. 2c (C2217), where we plot a different subset of the full dataset in order to bette show long-range strain gradients. The sharp jump in strain values seen 'Harp A' is due to small residues from the substrate placed at the clamping point, which lower the effective length of the bridge. Few points are missing because the corresponding bridges were not measurable because broke during the fabrication or ended up sticking to the substrate.

## VI.   DERIVATION OF FREQUENCY VS TEMPERATURE ANALYSIS EQUATIONS

**Derivation of the main text Eq.3**

The resonance frequencies of the flexural modes for a thin and long homogeneous cantilever having rectangular shape (thickness $t$, length $L$, width $w$) are given by

$$f_n = \frac{\lambda_n^2}{2\pi} \frac{t}{L^2} \sqrt{\frac{E}{12\rho(1-\nu^2)}}, \tag{SE 5}$$

where $n$ is the index of the mode number, $\rho$ is the mass density, $E$ is the Young's modulus, and $\nu$ the Poisson's ratio. We now write the squared frequency by considering the first flexural mode and by putting together the numerical factors

$$f_1^2 = k_1 \frac{t^2}{L^4} \frac{E}{\rho} \; ; \; k_1 = \frac{\lambda_1^4}{48\pi^2(1-\nu^2)} \tag{SE 6}$$

From this we can calculate the relative derivative with respect to the temperature

$$\frac{\partial_T f_1^2}{f_1^2} = \frac{k_1}{f_1^2} \left[ 2t\partial_T t \frac{E}{L^4\rho} - 4\partial_T L \frac{tE}{L^5\rho} - \partial_T \rho \frac{tE}{L^4\rho^2} + \partial_T E \frac{t}{L^4\rho} \right] =$$
$$= 2\frac{\partial_T t}{t} - 4\frac{\partial_T L}{L} - \frac{\partial_T \rho}{\rho} + \frac{\partial_T E}{E} \tag{SE 7}$$

the first and second terms correspond to the definition of the thermal expansion coefficient along the out/in-plane direction. Here, the thermal expansion is considered isotropic and temperature-dependent ($\alpha(T)$). By imposing constant mass, the density contribution becomes

$$-\frac{\partial_T \rho}{\rho} = -\frac{\partial_T(M/V)}{M/V} = -\partial_T \left( \frac{1}{V} \right) V = - \left( -\frac{\partial_T V}{V^2} V \right) = \frac{\partial_T V}{V} \tag{SE 8}$$

where $V$ is the volume. Under first order approximation, this lead to the Eq. 3 of the main text

$$\frac{\partial_T f_1^2}{f_1^2} = +2\alpha - 4\alpha + 3\alpha + \frac{\partial_T E}{E} = \alpha + \frac{\partial_T E}{E}. \tag{SE 9}$$

**Derivation of the main text Eq.4**

To obtain the expression for Eq. 4 reported in the main text, we start from Eq. SE 6 and consider the first-order approximation for the thermal expansion for each geometric

variable $(t, L, w)$

$$t = t_0(1 + A(T))$$
$$L = L_0(1 + A(T))$$
$$w = w_0(1 + A(T))$$
$$\rho = \frac{M}{V} = \frac{M}{L_0 t_0 w_0 * (1 + A(T))^3} = \rho_0(1 + A(T))^{-3}$$

where $A(T)$ is the integral function of the linear thermal expansion coefficient between $T_0$ and $T$. Here, $T_0 = 298$ K is the temperature at which we measured the Young's modulus from the length dependence of the first eigenfrequency.

$$A(T) = \int_{T_0}^{T} \alpha(T) dT$$

$A(T)$ is calculated by numerical integration of the literature data reported in Figure 4b of the main text. High-temperature values are obtained by constant extrapolation of measured ones. From the expressions above, we can re-write Eq. SE 6 as

$$
\begin{aligned}
f_1^2(T) =& k_1 \frac{t_0^2(1 + A(T))^2}{L_0^4(1 + A(T))^4} \frac{E(T)}{\rho_0(1 + A(T))^{-3}} \\
=& k_1 \frac{t_0^2}{L_0^4} \frac{E(T)}{\rho_0}(1 + A(T)) \\
=& f_{1,T_0}^2 \frac{E(T)}{E_{T_0}}(1 + A(T))
\end{aligned}
\tag{SE 10}
$$

The Eq. 4 in the main text is thus obtained by inverting the equation above

$$E(T) = E_{T_0} \frac{f_1^2(T)}{f_{1,T_0}^2} \frac{1}{(1 + A(T))} \tag{SE 11}$$

## VII. FREQUENCY VS TEMPERATURE DATA MEASURED ON DIFFERENT CANTILEVERS

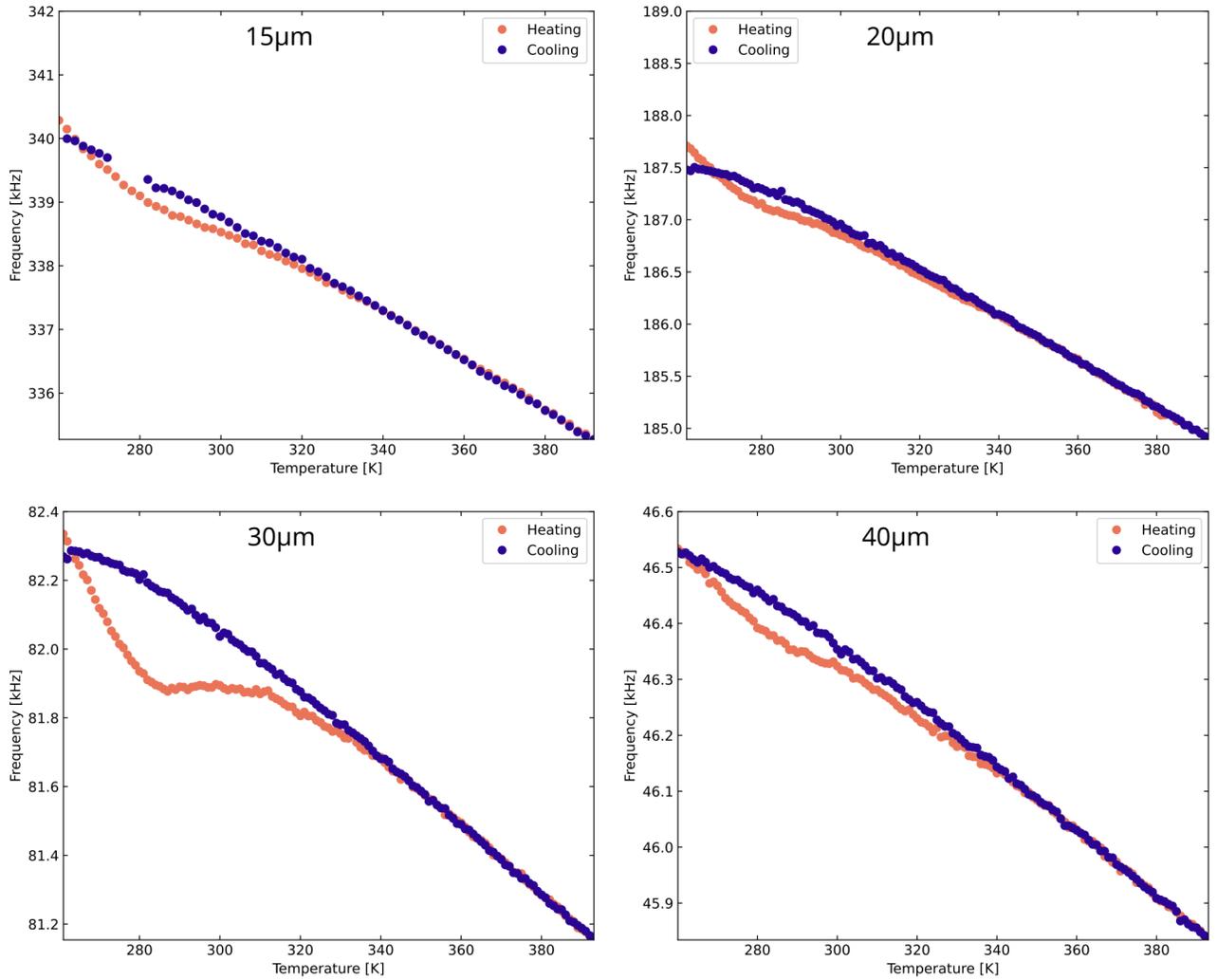

Figure S6. Eigenfrequency of the first flexural mode for cantilevers having different length, as indicated in each panel label. These measurements were acquired on C2203, while those reported in the main text on C2217. Thermal hysteresis is always visible, but with different magnitude. This because the closing of the hysteresis loop is located below the accessible span of our measurement setup. These data are thus nested cycles, whose loop size is critically dependent on the lowest temperature achieved at the beginning the data acquisition. Nominally, all the initial temperatures are in a 0.5 K range around 258 K. However we do not have direct information about local cantilever temperature, which may slightly differ from the thermometer.



\end{document}